\documentclass[10pt,conference]{IEEEtran}
\IEEEoverridecommandlockouts

\usepackage{cite}
\usepackage{amsmath,amssymb,amsfonts}

\usepackage{algorithmic}
\usepackage[ruled, linesnumbered, vlined]{algorithm2e} 

\usepackage{graphicx}
\usepackage{textcomp}
\usepackage{xcolor}
\graphicspath{ {./images/} }
\usepackage{booktabs}
\usepackage{lscape}
\usepackage{adjustbox}

\usepackage{caption}
\usepackage{subcaption}
\usepackage[left=1.62cm,right=1.62cm,top=0.8in]{geometry}
\usepackage{lipsum} 
\usepackage{flafter} 
\usepackage{color,soul} 
\usepackage{tabularx,booktabs} 
\newcolumntype{C}{>{\centering\arraybackslash}X} 
\usepackage[bookmarks=false,hidelinks,draft]{hyperref} 
\usepackage{scalerel} 
\usepackage{tikz} 

\renewcommand{\baselinestretch}{1.01}
\def\BibTeX{{\rm B\kern-.05em{\sc i\kern-.025em b}\kern-.08em T\kern-.1667em\lower.7ex\hbox{E}\kern-.125emX}}
\usepackage{stfloats}

\usetikzlibrary{svg.path}
\definecolor{orcidlogocol}{HTML}{A6CE39}
\tikzset{
	orcidlogo/.pic={
		\fill[orcidlogocol] svg{M256,128c0,70.7-57.3,128-128,128C57.3,256,0,198.7,0,128C0,57.3,57.3,0,128,0C198.7,0,256,57.3,256,128z};
		\fill[white] svg{M86.3,186.2H70.9V79.1h15.4v48.4V186.2z}
		svg{M108.9,79.1h41.6c39.6,0,57,28.3,57,53.6c0,27.5-21.5,53.6-56.8,53.6h-41.8V79.1z M124.3,172.4h24.5c34.9,0,42.9-26.5,42.9-39.7c0-21.5-13.7-39.7-43.7-39.7h-23.7V172.4z}
		svg{M88.7,56.8c0,5.5-4.5,10.1-10.1,10.1c-5.6,0-10.1-4.6-10.1-10.1c0-5.6,4.5-10.1,10.1-10.1C84.2,46.7,88.7,51.3,88.7,56.8z};
	}
}
\newcommand{\orcidicon}[1]{\href{https://orcid.org/#1}{\mbox{\scalerel*{
				\begin{tikzpicture}[yscale=-1,transform shape]
				\pic{orcidlogo};
				\end{tikzpicture}
			}{|}}}}

\newcommand{\ignore}[1]{}

\newcommand{\linebreakand}{%
\end{@IEEEauthorhalign}
\hfill\mbox{}\par
\mbox{}\hfill\hspace*{-1cm}\begin{@IEEEauthorhalign} 
}

\begin{document}

\title{Integrated 5G mmWave Positioning in Deep Urban Environments: Advantages and Challenges}

\author{ 
Sharief Saleh\textsuperscript{\orcidicon{0000-0003-1365-417X}}$^{* \dag}$, Qamar Bader\textsuperscript{\orcidicon{0000-0002-4667-1710}}$^{* \dag}$,
Malek~Karaim\textsuperscript{\orcidicon{0000-0002-2897-1330}}$^{\dag}$,
Mohamed Elhabiby\textsuperscript{\orcidicon{0000-0002-1909-7506}}$^{\ddag}$,
and Aboelmagd Noureldin\textsuperscript{\orcidicon{0000-0001-6614-7783}}$^{*\dag}$\\

\IEEEauthorblockA{
$^*$ Department of Electrical and Computer Engineering, Queen's University, Kingston, ON, Canada\\
$^\dag$ Navigation and Instrumentation (NavINST) Research Lab,\\ Department of Electrical and Computer Engineering, Royal Military College of Canada, Kingston, ON, Canada\\
$^{\ddag}$Micro Engineering Tech. Inc.\\
Email: \{sharief.saleh, qamar.bader, nourelda\}@queensu.ca}

}
\maketitle

\begin{abstract}	
Achieving the highest levels of autonomy within autonomous vehicles (AVs) requires a precise and dependable positioning solution that is not influenced by the environment. 5G mmWave signals have been extensively studied in the literature to provide such a positioning solution. Yet, it is evident that 5G alone will not be able to provide uninterrupted positioning services, as outages are inevitable to occur. Towards that end, few works have explored the benefits of integrating mmWave positioning with onboard motion sensors (OBMS) like inertial measurement units (IMUs) and odometers. Inspired by INS-GNSS integration literature, all methods defaulted to a tightly-coupled (TC) integration scheme, which hinders the potential of such an integration. Additionally, the proposed methods were validated using simulated 5G and INS data with probability-based line-of-sight (LOS) assumptions. Such an experimental setup fails to highlight the true advantages and challenges of 5G-OBMS integration.  Therefore, this study first explores a loosely-coupled (LC) 5G-OBMS integration scheme as a viable alternative to TC schemes. Next, it examines the merits and challenges of such an integration in a deep-urban setting using a novel quasi-real simulation setup. The setup comprises quasi-real 5G measurements from the Siradel simulator and real commercial-grade IMU measurements from a challenging one-hour-long trajectory in downtown Toronto. The trajectory featured multiple natural 5G outages which helped with assessing the integration's performance. The proposed LC method achieved a $14$-cm level of accuracy for $95\%$ of the time, while significantly limiting positioning errors during natural 5G outages.
\end{abstract}


\section{Introduction}
The potential benefits of achieving high levels of autonomy in autonomous vehicles (AVs) are numerous, including increased safety, enhanced reliability, and decreased congestion rates, which will significantly improve the transportation network of cities \cite{AVCongestion}. Hence, it is expected that the North American AVs market will grow at a compound annual growth rate (CAGR) of $17.1\%$ over the next years, reaching a value of $52.3\$B$ by 2030 \cite{CAGR}. One of the main prerequisites towards full driving autonomy is the availability of an uninterrupted positioning solution at a decimeter level of accuracy for $95\%$ of the time.

Towards that end, various positioning-capable technologies such as global navigation satellite systems (GNSS), network-based receivers, inertial navigation systems (INS), odometers, and perception-based systems are currently installed in modern vehicles. Researchers have been investigating and developing these positioning technologies for decades to meet the demands of the AV market. However, no single positioning technology can provide a reliable solution for all possible driving dynamics in all environments. GNSS, for example, may provide centimeter-level accuracy in open sky environments, but their performance deteriorates significantly in urban canyons due to signal multipath and blockage  \cite{ProfBook}. Alternative wireless positioning solutions like WiFi and ultra-wideband (UWB) could be utilized in urban canyon scenarios. However, they require a dense deployment that may be prohibitive due to their short range of communication \cite{Koivisto2017}. Perception-based systems, such as cameras, LiDARs, and radars, may work in urban canyon scenarios but they suffer from performance degradation due to partially overshadowed scenes or weather conditions \cite{Camera,Radar}. Dead-reckoning-based systems like inertial measurement units (IMUs) and odometers; collectively known as onboard motion sensors (OBMS), do not suffer from signal blockage but they accumulate errors due to their inherent design \cite{ProfBook}.

The emerging 5G new radio (NR) cellular network, on the other hand, is equipped with key technologies that can support high-precision positioning in GNSS-denied environments like urban canyons. Moreover, unlike UWB and WiFi technologies, 5G has a much longer communication range and will be densely deployed in urban canyons for communications purposes. Additionally, 5G NR is weather independent, unlike perception-based systems. Furthermore, 5G positioning signals will have a bandwidth of $400$ MHz, resulting in notably accurate time-based measurements such as time of arrival (TOA), round-trip time (RTT), and time difference of arrival (TDOA) \cite{wymeersch_radio_2022}. In addition, 5G systems will be equipped with massive MIMO capabilities, enabling precise downlink angle of departure (DL-AOD) and uplink angle of arrival (UL-AOA) measurements \cite{merits}. Thus, enabling hybrid range and angle positioning solutions that can operate using a single base station (BS); unlike trilateration and triangulation-based methods which require at least three BSs to operate. Therefore, 5G NR is anticipated to take the role of GNSS satellites in urban canyon environments, providing position-fixing services to estimate and reset OBMS biases when available. Likewise, during times of 5G outage due to non-line of sight (NLOS) communications, OBMS is expected to bridge the gaps to ensure a seamless positioning solution. Hence, The integration of 5G mmWave positioning technology with OBMS has the potential to provide precise, continuous, and environment-independent positioning solutions for AVs in deep-urban environments where other positioning technologies may fail.

The integration of 5G with external sensors is a relatively unexplored research area, and only a few studies have addressed this problem. Authors in \cite{mostafavi_vehicular_2020} have proposed a tightly-coupled (TC) approach that directly integrates the 5G's AOD and TOA measurements with two accelerometers that are mounted on the x and y axes. Using accelerometers alone without incorporating gyroscope measurements is unusual, given that these sensors are typically used in conjunction with each other. Additionally, the authors utilized a constant acceleration transition model despite the availability of accelerometer measurements, which is odd. The authors conducted their simulations in 2D, which is not practical in real-world applications since it ignores the estimation of pitch and roll angles. Authors in \cite{luo_research_2021} proposed the use of an invariant extended Kalman filter (InEKF) to integrate horizontal-AOD, vertical-AOD, and TOA measurements in a TC fashion. The proposed method employs six IMU sensors for the state transition process, providing an advantage over the prior work's constant acceleration model. The filter's states include position, velocity, and attitude (PVA) navigation states, along with the six IMU biases, which is a conventional practice in Global Navigation Satellite System (GNSS)-Inertial Navigation System (INS) filters. The authors validated the proposed approach using a fully simulated environment for both 5G and IMU measurements. In \cite{wang_simulation_2022}, the authors proposed a federated EKF approach that integrates measurements from 5G, INS, GNSS, and LEO satellites using hybrid TC and loosely coupled (LC) schemes. This method employs three sub-filters in addition to a main filter, where each sub-filter conducts TC integration between the IMU measurements and one of the three aiding technologies. The output of each sub-filter is then integrated within the main filter using an LC approach. The 5G-INS portion of the method utilizes TOA measurements from four BSs to conduct centralized trilateration, which may result in high linearization errors due to the close proximity of the UE to one of the connected BSs as discussed in \cite{DTCM,saleh_5g-enabled_2022}. It is also unrealistic to have access to four BSs in real-life scenarios. Additionally, the proposed method employs a linearized mechanization-based state transition model, which induces unnecessary linearization errors. Finally, results were generated using simulated 5G and IMU data.

To conclude, it is evident that there is a limited number of studies on the integration of 5G and OBMS. The existing approaches employ TC methods that result in linearization errors. Additionally, their validation relies solely on simulated data for 5G and IMU measurements, which may not fully represent the complexities present in real-world scenarios. Therefore, this work first introduces LC integration as an alternative to TC schemes. Second, this paper studies the advantages and challenges of integrating 5G with low-cost OMBS by utilizing a novel quasi-real simulation setup. The setup employs a mix of real IMU data from an hour-long trajectory in downtown Toronto along with high-fidelity simulation of 5G signals generated by the Siradel simulator.

The paper is structured in the following manner: In Section II, the system model is established. Section III presents the 5G-OBMS integration methodology. The experimental setup used in the study is presented in Section IV, along with the findings and discussions around 5G-OBMS integration. Finally, Section V concludes the paper.

\section{System Model}
\subsection{System States}
In this work, the states of the Kalman filter comprise the position, velocity, and attitude states as well as the biases of the six gyroscopes and accelerometers. The 3D position of the UE is denoted as $(x,y,z)$ in the universal transverse Mercator (UTM) coordinate system and as $(\varphi,\lambda,h)$ in the world geodetic system (WGS). The velocity states of the vehicle represented in the local-level frame (l-frame) are expressed as $(v_e, v_n, v_u)$, which denotes the velocities in the east, north, and up directions, respectively. The vehicle velocities denoted in the body frame (b-frame) of the vehicle are expressed as $(v_x,v_y,v_z)$, which correspond to velocities in the longitudinal, lateral, and vertical directions, respectively. The angular rotations between the vehicle's b-frame and the l-frame constitute the attitude state of the vehicle. The attitude angles are the pitch, roll, and azimuth angles, which are expressed as $(p,r,a)$, respectively. The bias states of the three gyroscopes and the three accelerometers are $(\delta\omega_{x}, \delta\omega_{y}, \delta\omega_{z})$ and  $(\delta f_{x}, \delta f_{y}, \delta f_{z})$, respectively.

\subsection{5G Measurables}
The position of the 5G BSs is denoted as $(x_{BS},y_{BS},z_{BS})$. The range measurements between the UE and the BS can thus be formulated as seen in (\ref{Range}).

\begin{equation}\label{Range}
    r=\sqrt{\Delta x^2+ \Delta y^2 + \Delta z^2}
\end{equation}

\noindent Where $(\Delta x, \Delta y, \Delta z)$ denote the difference between the UE's position $(x,y,z)$ and the wireless node's position $(x_{BS},y_{BS},z_{BS})$. In this work, RTT range measurements are utilized:

\begin{equation}\label{RTT}
    \begin{split}
        &\tau_{RTT}=2\Delta t\\
        &r=c\frac{\tau_{RTT}}{2},
    \end{split}
\end{equation}

\noindent where $\Delta t$ is the time delay between the time of departure (TOD) recorded at the transmitter and the TOA measured at the receiver, and $c$ is the speed of light. The relative horizontal and vertical angles between the BS and the UE with respect to the BS's orientation can be formulated as shown in (\ref{Angle}).

\begin{equation}\label{Angle}
    \begin{split}
        \theta&=\tan^-1\left(\frac{\Delta y}{\Delta x}\right)\\
        \psi&=\sin^-1\left(\frac{\Delta z}{r}\right)
    \end{split}
\end{equation}

\noindent Where $\theta$ is the horizontal AOD and $\psi$ is the vertical AOD.

\subsection{OBMS Measurables}
IMUs typically comprise three orthogonal accelerometers and three orthogonal gyroscopes. The acceleration of the vehicle in the b-frame, $f_b=[f_x, f_y, f_z]$, can then be computed from the sensed forces of the accelerometers. Additionally, gyroscopes measure the rate of rotation of the vehicle around the axis of the gyroscope, $\omega_b=[\omega_x, \omega_y, \omega_z]$. Wheel odometer sensors, on the other hand, are used with wheel-based vehicles to compute the forward velocity of the vehicle, in the b-frame \cite{ProfBook}. This is done by measuring the number of turns the wheel turned per unit time; $\omega_{Odo}$. The resulting forward velocity is computed as follows \cite{ProfBook}: 

\begin{equation}\label{odo}
        v_{Odo}= 2\pi r_{wheel} \cdot \omega_{Odo},
\end{equation}

\noindent where $r_{wheel}$ is the radius of the wheel.

\section{Methodology}
\begin{figure}[t!]
	\centering
	\includegraphics[width=\columnwidth,trim=15 18 15 15,clip]{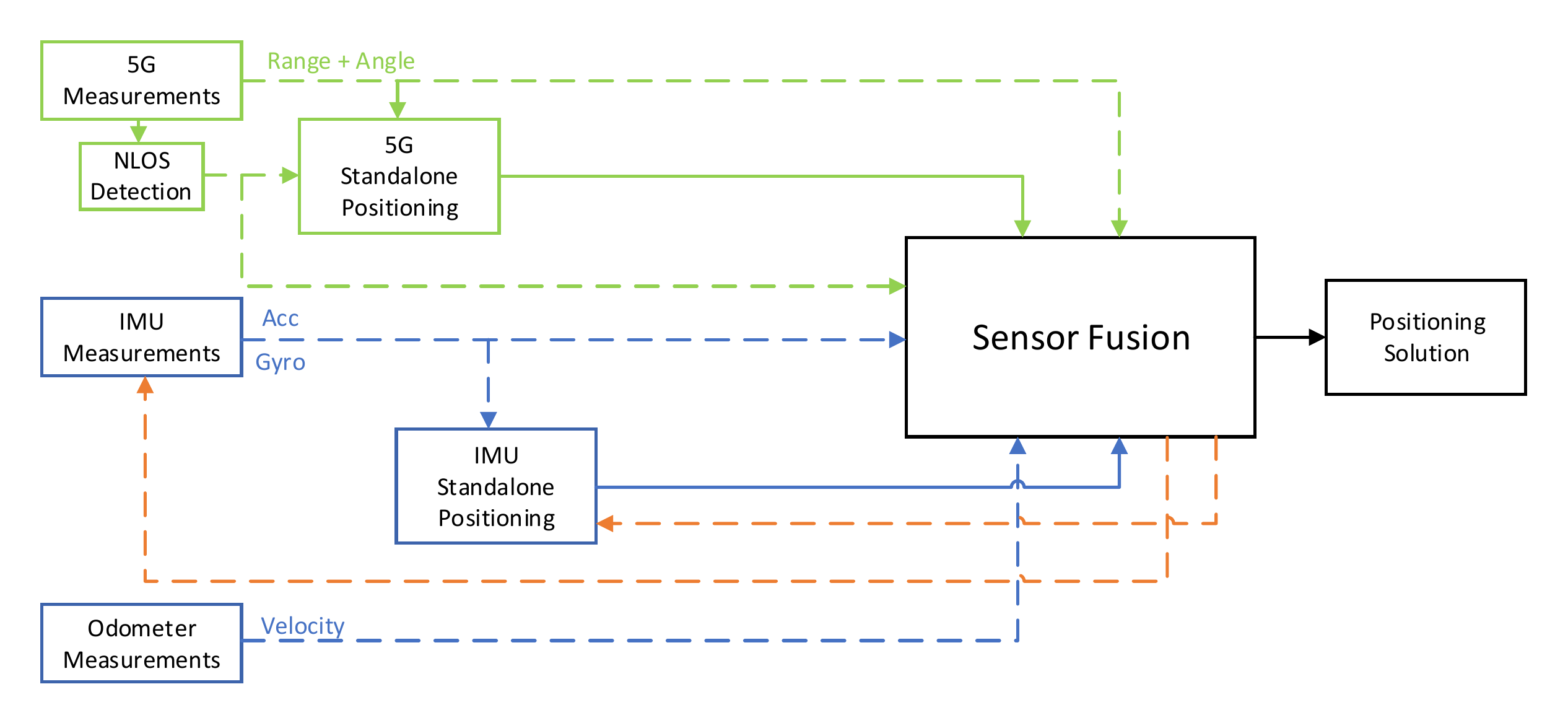}
	\DeclareGraphicsExtensions.
	\caption{Proposed LC integration between 5G and OBMS.}
	\label{Methodology}
\end{figure}

The overall positioning methodology consists of three stages as shown in Fig. \ref{Methodology}. First, a 5G standalone line of sight (LOS) positioning solution is computed with the aid of an NLOS detection scheme, proposed in \cite{bader_nlos_2022}, and an LC KF integration scheme proposed in \cite{saleh_would_2022}. Second, a dead-reckoning-based solution is computed via the mechanization of the accelerometer and gyroscope measurements. Finally, both positioning solutions are fused in addition to odometer measurements in an LC EKF to provide an enhanced positioning solution.

\subsection{5G Standalone Positioning}
In this work, we consider that BSs have access to a uniform linear array (ULA) of antennas, enabling horizontal AOD measurements only. To compute the 3D position of the UE, an assumption should be made about the height of the UE. That is, we assume that the UE maintains a constant height, which is a fair assumption for AVs. Therefore, the position of the UE can be computed as shown in (\ref{2D Pos}).

\begin{equation}\label{2D Pos}
    \begin{split}
        &x=r_{2D} \cdot \sin(\theta) +x_{BS}\\
        &y=r_{2D} \cdot \cos(\theta) +y_{BS}\\
        &r_{2D}=\sqrt{r^2-\Delta z^2}
    \end{split}
\end{equation}

\noindent Where $r_{2D}$ is the 2D range between the BS and the UE. Such computation is done for each participating BS. In order to choose the participating LOS BSs, an NLOS detection algorithm, proposed in \cite{bader_nlos_2022}, is utilized. The method relies on the difference between the power-based range measurement and the time-based range measurement to differentiate between LOS and NLOS signals. The selected BSs are then fused in an LC fashion within a linear KF. The filter utilizes a constant velocity transition model. The details of the filter design can be found in \cite{saleh_would_2022}. The output of the filter would be a standalone 5G LOS positioning solution.

\subsection{INS Mechanization}
\begin{figure}[t!]
	\centering
	\includegraphics[width=\columnwidth]{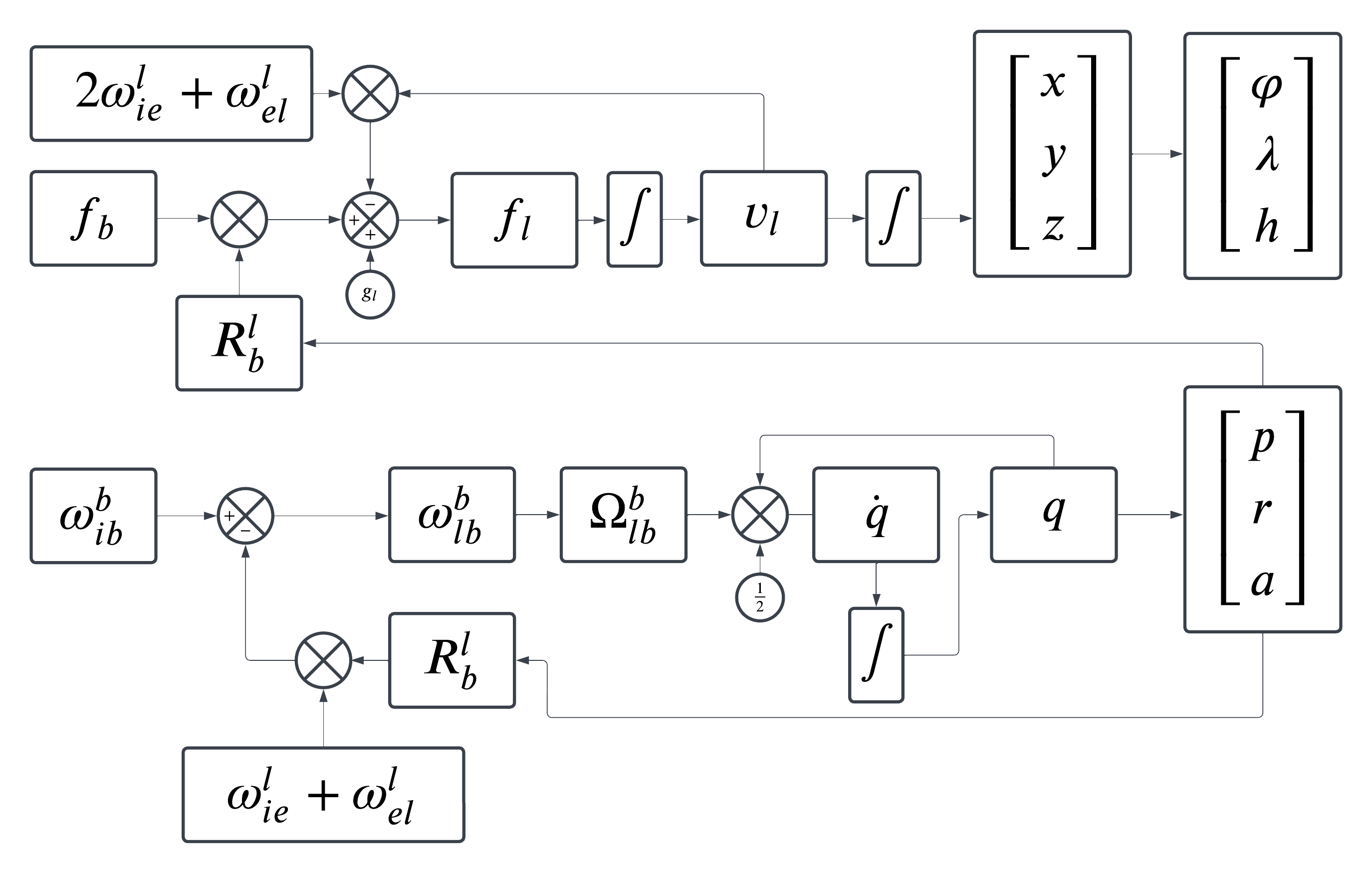}
	\DeclareGraphicsExtensions.
	\caption{INS mechanization block diagram.}
	\label{INS Mech}
\end{figure}

The process of computing the PVA states through the accumulation and transformation of acceleration and gyroscope measurements is referred to as INS mechanization. Figure \ref{INS Mech} illustrates the general block diagram for INS mechanization, as described in \cite{ProfBook}. The INS mechanization process is composed of two main steps. The first step involves computing the attitude angles by eliminating external effects, projecting, and accumulating gyroscope measurements. The second step involves computing velocities and positions in the l-frame by eliminating extrinsic  effects, projecting, and accumulating accelerometer measurements.

Both gyroscope and accelerometer measurements are affected by two external factors, which are the alteration in orientation caused by motion on non-linear coordinates and the rotational speed of the Earth relative to the inertial frame. These effects are captured by $\boldsymbol{\omega}_{el}^l$ and $\boldsymbol{\omega}_{ie}^l$, respectively \cite{ProfBook}. An additional external factor that should be compensated in accelerometer measurements is the effect of the earth's gravity on the three sensors, denoted as $\boldsymbol{g}_l$ \cite{ProfBook}. The rotation matrix $\boldsymbol{R}_b^l$ is used to transform states and measurements from the b-frame to the l-frame. The transformation to the l-frame is important in order to have a meaningful positioning solution that can be used in conjunction with maps \cite{ProfBook}. The final step of the mechanization process would be the integration/accumulation of the measurables to compute the position, velocity, and attitude states. In order to compute the velocity of the vehicle in the l-frame, a single integration of the l-frame acceleration is needed. These velocities are integrated a second time to compute the position of the vehicle. To compute the attitude states, however, quaternions are utilized, which are denoted as $\boldsymbol{q}$, \cite{ProfBook}. Quaternions are a mathematical concept used to represent the orientation of an object in 3D space, allowing for more efficient and accurate computations than traditional methods \cite{ProfBook}. The computation of the quaternion vector in the current epoch $k$ relies on the previous epoch's quaternion $\boldsymbol{q}_{k-1}$ and the current epoch's gyroscope measurements' skew matrix $\boldsymbol{\Omega}_{lb_k}^b$, as shown in (\ref{qdot}). The quaternion vector can then be easily transformed back to the pitch, roll, and azimuth angles as detailed in \cite{ProfBook}.

\begin{equation}\label{qdot}
    \boldsymbol{q}_{k}= \boldsymbol{q}_{k-1} +  \frac{1}{2} \Delta t \boldsymbol{\Omega}_{lb_k}^b \boldsymbol{q}_{k-1}
\end{equation}

\subsection{Sensor fusion}
The integration between the 5G and OBMS technologies utilizes an EKF in an LC fusion scheme. The states of the filter are shown in (\ref{States}).

\begin{equation}\label{States}
    \boldsymbol{x}=\begin{bmatrix}
        &\varphi &\lambda &h &\dots\\
        &v_e &v_n &v_u &\dots\\
        &p &r &a &\dots\\
        &\delta\omega_{x} &\delta\omega_{y} &\delta\omega_{z} &\dots\\
        &\delta f_{x} &\delta f_{y} &\delta f_{z}
    \end{bmatrix}^T
\end{equation}

The transition model of the first nine PVA states comprises the mechanization process described earlier. The transition model of the six sensor biases is characterized by the first-order Gauss-Markov model shown in (\ref{Gauss}).

\begin{equation}\label{Gauss}
\begin{split}
        \delta \Dot{\omega}&=-\beta_\omega \delta \omega + \sqrt{2\beta_\omega}w_{\omega_B}(t)\\
        \delta \Dot{f}&=-\beta_f \delta f + \sqrt{2\beta_f}w_{f_B}(t)
\end{split}
\end{equation}

The measurement vector of the filter, shown in (\ref{Z}), comprises the fused 5G standalone positioning solution and the odometer velocity projected in the l-frame.

\begin{equation}\label{Z}
    \boldsymbol{z}= \begin{bmatrix}
        \varphi_{5G} & \lambda_{5G} & h_{5G} &  v_{e_{Odo}} & v_{n_{Odo}} & v_{u_{Odo}}
    \end{bmatrix}^T
\end{equation}

\noindent It can be clearly seen that the relationship between the first six states and measurements is indeed linear. Hence, no linearization errors will be endured in the proposed LC method; as opposed to the TC methods utilized in the literature.

\section{Experimental Setup and Results}
\subsection{Experimental Setup}
\begin{figure}[t!]
	\centering
	\includegraphics[width=\columnwidth]{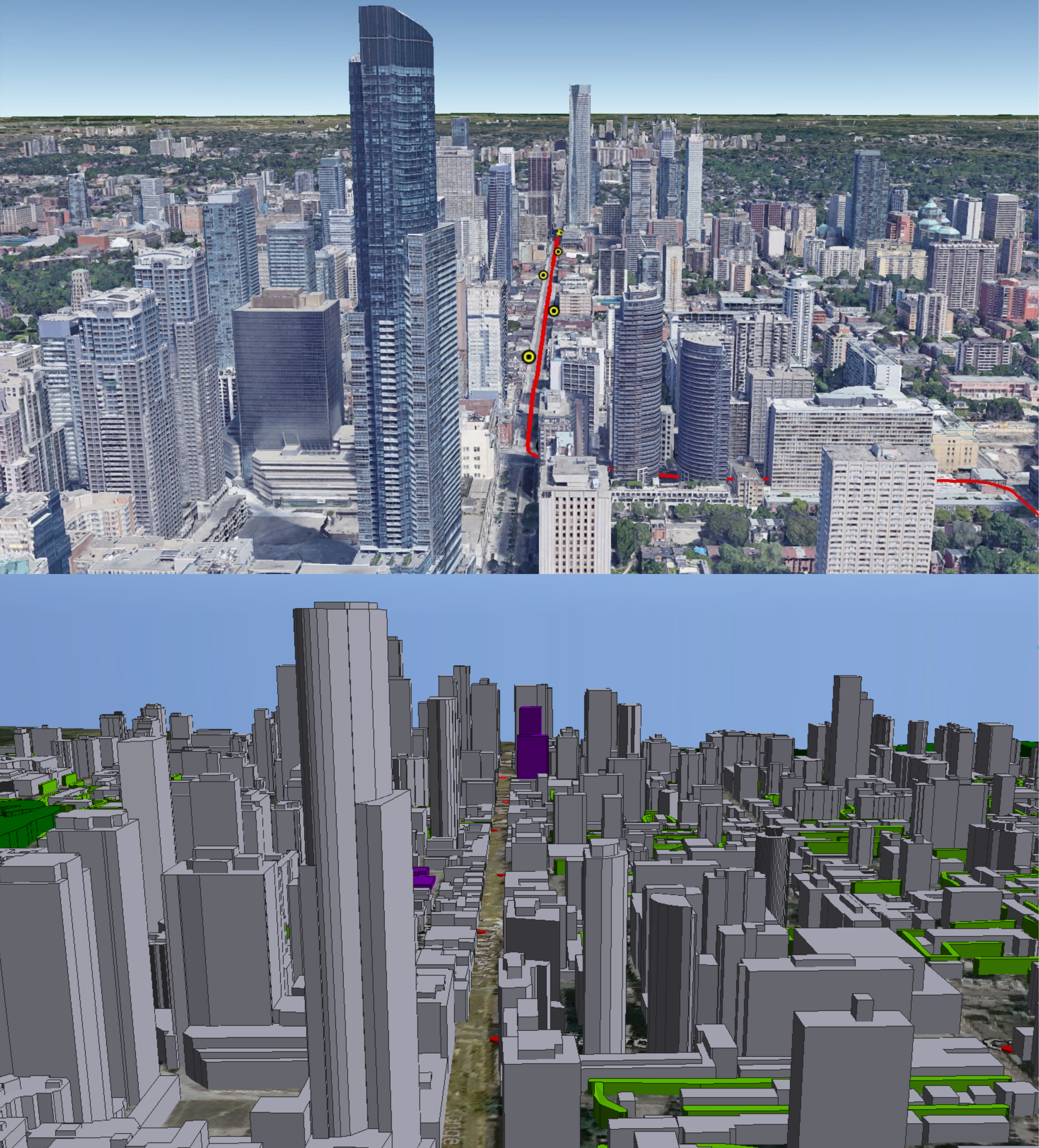}
	\DeclareGraphicsExtensions.
	\caption{Downtown Toronto, ON, Google Earth (top) vs. Siradel simulation tool (bottom).}
	\label{GoogleEarth vs Siradel}
\end{figure}

\begin{figure}[t!]
	\centering
	\includegraphics[width=\columnwidth]{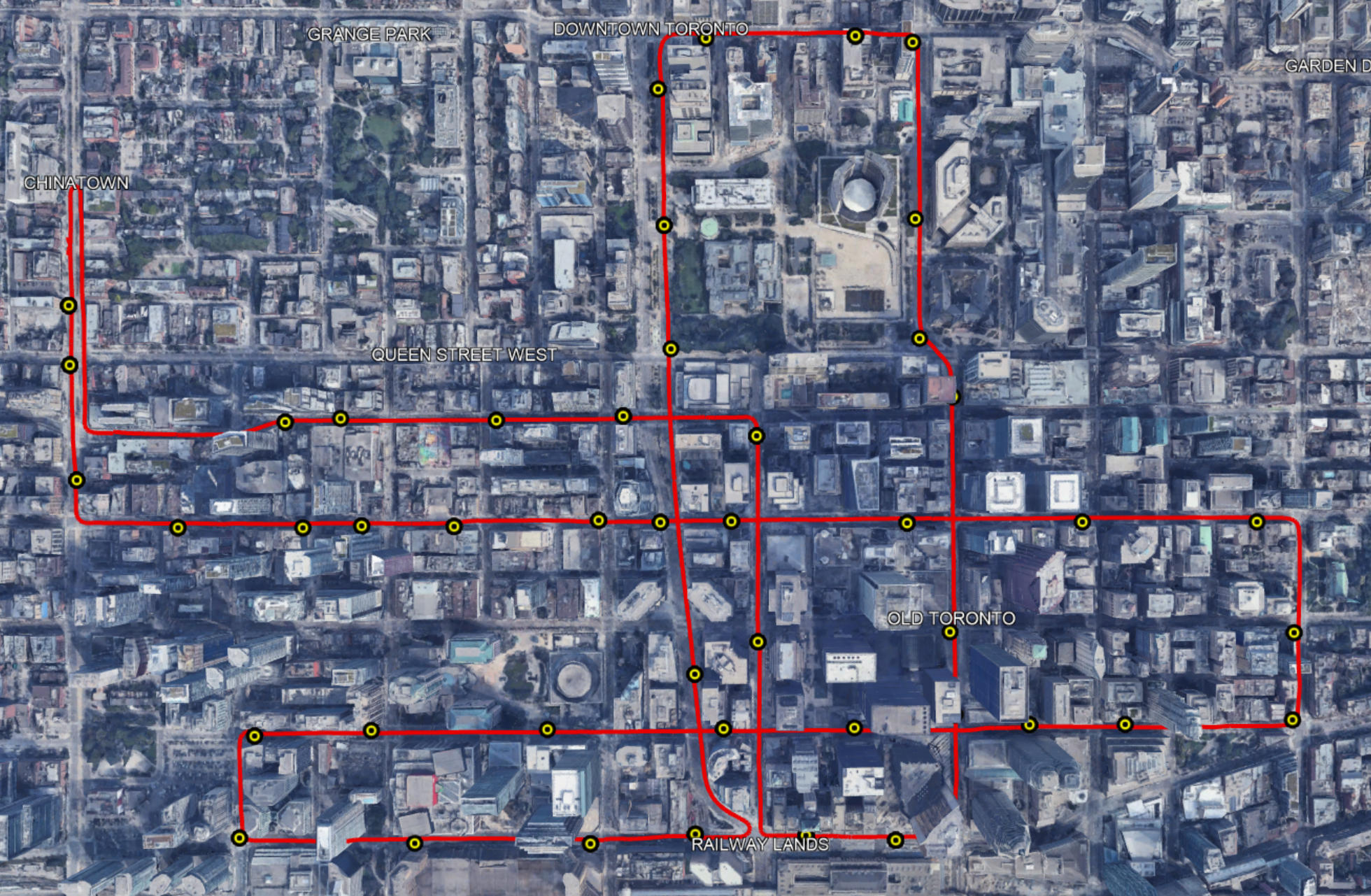}
	\DeclareGraphicsExtensions.
	\caption{Downtown Toronto Trajectory.}
	\label{Traj}
\end{figure}

In order to assess the merits and challenges of integrated 5G positioning, a quasi-real simulation setup was created using Siradel's 5G\_Channel simulator. Siradel comprises LiDAR-based maps of the buildings, vegetation, and water bodies in downtown areas of dense cities like Toronto, Canada, as seen in Fig. \ref{GoogleEarth vs Siradel}. The simulator requires the position of the UE and the virtually connected BSs to compute the required positioning measurables. To mimic a real-life trajectory, a vehicle equipped with a high-end positioning solution from NovAtel's ProPak6 was driven in the deep-urban area of downtown Toronto, generating the ground truth trajectory. The trajectory, shown in Fig. \ref{Traj}, is $1$ hr and $13$ mins long and traversed a distance of $9$ km. Next, an algorithm was developed to deploy BSs along the driven trajectory, with $250$ m inter-cell distance; according to 3GPP standards \cite{Course}. The BSs were configured within Siradel to operate at mmWave $28$ GHz frequency and a bandwidth of $400$ MHz. Finally, the generated BS positions and the reference solution were imported to Siradel to generate the 5G measurables. In addition to that, the vehicle was equipped with a commercial-grade IMU from TPI and an odometer from OBD-II to provide real data for integration purposes. 



\subsection{Results and Discussions}
\begin{figure}[t!]
	\centering
	\includegraphics[trim=117pt 240 125pt 245pt,clip,width=\columnwidth]{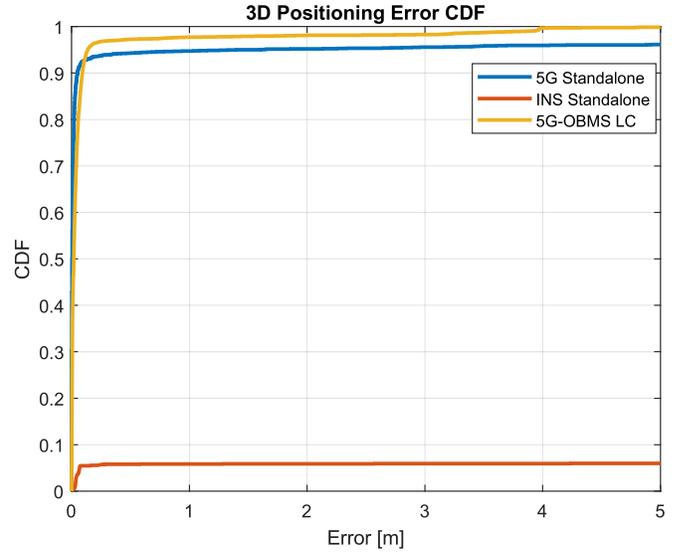}
	\DeclareGraphicsExtensions.
	\caption{CDF of 3D positioning errors.}
	\label{Integration CDF}
\end{figure}

The simulation of the hour-long trajectory revealed that a UE could be linked to a maximum of three BSs simultaneously, with a likelihood of $30\%$. The analysis further indicated that the UE would be able to connect to two BSs for $46\%$ of the time, and to a single BS for only $21\%$ of the time. Additionally, there is a low probability, $3\%$, of experiencing a complete loss of LOS connection to all BSs. In fact, four natural 5G outages were encountered during the course of the trajectory with varying degrees of severity. To assess the importance of 5G-OBMS integration, the proposed LC integration scheme is compared to the standalone 5G and INS solutions that constitute the integrated solution. The resulting CDF of the 3D positioning error is depicted in Fig. \ref{Integration CDF}.
\noindent As anticipated, the commercial-grade INS standalone solution was not able to provide a stable positioning solution for a long period of time. On the other hand, it can be seen that there is virtually no difference between the standalone 5G solution and the integrated 5G-OBMS solution in the centimeter error range, as they both performed relatively well. This is expected as low-cost IMUs cannot provide high-accurate positioning support for the rather precise 5G mmWave positioning solution. Yet, at higher error ranges, the 5G-OBMS solution significantly outperforms the standalone 5G solution. This is mainly due to the fact that high 5G positioning errors are caused by total 5G outages, where the standalone 5G multi-BS fusion filter relies on a constant velocity model. On the other hand, integrated 5G-OBMS seamlessly bridges such gaps by solely relying on the IMU and the odometer. Table \ref{INS vs 5G} shows a summary of the positioning error statistics of the three solutions. It can be seen that the standalone INS solution was not able to withstand a sub-meter level of accuracy for more than $6\%$ of the time. This is mainly due to the quadratic and cubic accumulation of accelerometer and gyroscope biases, respectively, over a long period of time. Such biases should be estimated and mitigated by the integration filter. Table \ref{INS vs 5G} also shows that the integrated 5G-OBMS has significantly outperformed the standalone 5G solution in terms of maximum and RMS errors. Moreover, minor enhancements are noticed for sub-meter and sub-decimeter error statistics for the aforementioned reasons. Finally, the proposed integrated solution was able to achieve an accuracy of $14$ cm for $95\%$ of the time.

\begin{table}[b!]
	\caption{3D Positioning Error Statistics Summary}
	\label{INS vs 5G}
	\begin{tabularx}{\columnwidth}{@{}l*{3}{C}c@{}}
		\toprule
		&Statistics &5G-SA 	    &INS-SA    &5G-OBMS\\
		\midrule
		&RMS         & 9 m      &1e4 m     &0.5 m\\ 
		&Max         & 89.3 m   &1.7e4 m   &6.3 m\\
		&$<2$ m      & 95.2\% 	&5.9\%     &98.1\%\\ 
		&$<1$ m      & 94.7\% 	&5.8\%     &97.7\%\\
		&$<30$ cm    & 93.9\% 	&5.7\%     &96.9\%\\
		\bottomrule
	\end{tabularx}
\end{table}

\subsection{Merits and Challenges of 5G-OBMS Integration}
The discussion above shows that there are two merits to integrating 5G and OBMS solutions. First, the integrated solution adequately and seamlessly bridges inevitable 5G outages, as seen in the results. This, in turn, greatly bounds the maximum positioning error to a tolerable range. Second, the integrated solution provides the IMU with numerous error resetting opportunities in GNSS-denied environments, like urban canyons. Hence, the IMU will have the ability to start a dead-reckoning solution during 5G outages with bias-free measurements and states.

Yet, as the 5G-OBMS integration research community is at its early stages of evolving, there is a myriad of challenges to be addressed. First, proper tuning of the measurement and process covariance matrices of the EKF is needed in order to accurately estimate the biases, which is a great challenge. Second, the authors believe that the standalone INS solution can be enhanced further by pausing the mechanization process while the vehicle is stationary. This will also halt the accumulation of residual sensor biases if any. Finding an adequate stationarity detection mechanism will pose a challenge. Finally, it was observed that the aiding 5G position measurements had weak observability on the attitude and gyroscope bias states. Meaning, having accurate position measurements will not guarantee proper estimation and correction of errors in the said states. In order to solve such a challenge, the 5G solution should be able to provide attitude-based corrections, which is also known as 6D state estimation. To achieve that, the UE must have access to a 2D antenna array in order to measure DL-AOA, which imposes an economical challenge.

\section{Conclusion}
Although 5G NR promises high BS densification, outages are expected to occur. According  to the study, absolute 5G outages will happen for $3\%$ of the time. Despite its low probability of occurrence, outages are the main source of errors and they cause intolerable amounts of positioning errors. It was shown in this work that the integration between 5G and OBMS has the capability to smoothly bridge such outages and maintain a low-error profile. It was also shown that 5G can indeed replace GNSS in urban areas and provide the INS with the needed bias estimation and correction services. Moreover, it was shown that such integration comes with a set of challenges that needs to be addressed. Namely, the tuning of the fusion filter, the mechanization stopping mechanism, and the observability of 5G measurables over the filter's states. Finally, an LC EKF-based fusion filter was proposed as an alternative to TC schemes to avoid linearization errors. The proposed LC method has achieved a $14$ cm level of accuracy for $95\%$ of the time while significantly bounding the positioning errors during natural 5G outages. All of the analysis and methods were accomplished using a novel setup comprising quasi-real 5G measurements from Siradel and real IMU measurements from a challenging hour-long trajectory in downtown Toronto.

\bibliographystyle{IEEEtran}
\bibliography{References,references}
\end{document}